\documentclass[10pt,aps,prd,preprintnumbers,showpacs,twocolumn,nofootinbib,notitlepage,longbibliography,superscriptaddress]{revtex4-1}

\usepackage{amsmath, amssymb,braket}
\usepackage{color}
\usepackage{float}
\usepackage{graphicx}
\usepackage{subfigure}
\usepackage{natbib}
\usepackage{adjustbox}
\usepackage{tikz}
\usetikzlibrary{shapes,arrows}
\usepackage[compat=1.1.0]{tikz-feynman}
\usepackage[colorlinks=true,citecolor=blue,urlcolor=blue]{hyperref}

\newcommand\orcid[1]{\href{http://orcid.org/#1}{\adjustbox{trim={-.15\width} {0\height} {-.15\width} {0\height},clip}{\includegraphics[height=10pt]{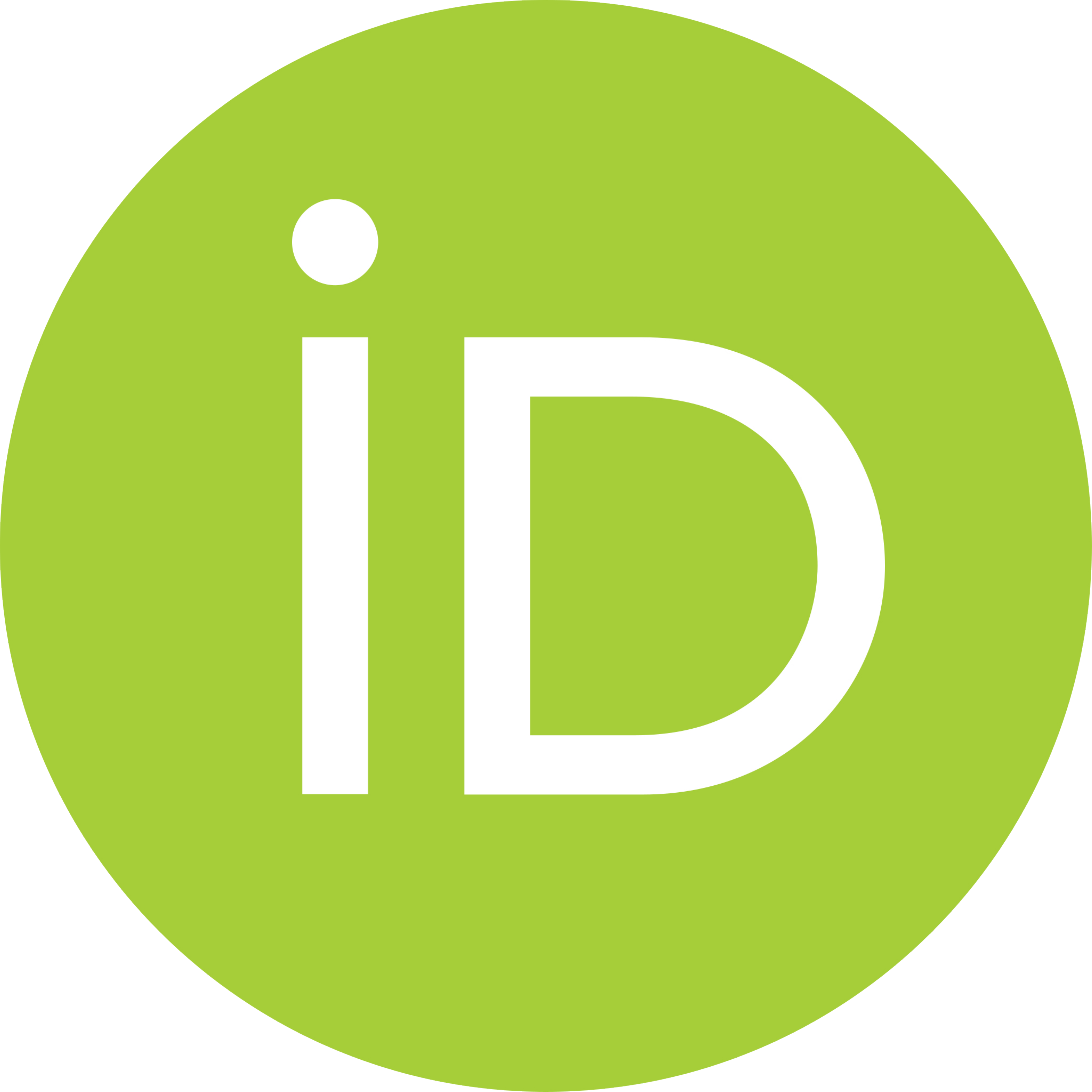}}}}
\newcommand{\n}{\nonumber \\}
\newcommand{\dbar}{d\hspace*{-0.08em}\bar{}\hspace*{0.1em}}

\begin{document}
\title{Effective bias expansion for circumventing 21 cm foregrounds}

\author{Wenzer Qin\orcid{0000-0001-7849-6585}}
\affiliation{Department of Physics, Massachusetts Institute of Technology, Cambridge, MA 02139, USA}
\affiliation{Center for Theoretical Physics, Massachusetts Institute of Technology, Cambridge, Massachusetts 02139, USA}
\affiliation{Center for Cosmology and Particle Physics, Department of Physics, New York University, New York, NY 10003, USA}

\author{Kai-Feng Chen\orcid{0000-0002-3839-0230}}
\affiliation{Department of Physics, Massachusetts Institute of Technology, Cambridge, MA 02139, USA}
\affiliation{MIT Kavli Institute, Massachusetts Institute of Technology, Cambridge, MA 02139, USA}

\author{Katelin Schutz\orcid{0000-0003-4812-5358}}
\affiliation{Department of Physics \& Trottier Space Institute, McGill University, Montr\'{e}al, QC H3A 2T8, Canada}

\author{Adrian Liu\orcid{0000-0001-6876-0928}}
\affiliation{Department of Physics \& Trottier Space Institute, McGill University, Montr\'{e}al, QC H3A 2T8, Canada}

\begin{abstract}
    \noindent
    The 21\,cm line of neutral hydrogen is a promising probe of the Epoch of Reionization (EoR) but suffers from contamination by spectrally smooth foregrounds, which obscure the large-scale signal along the line of sight. We explore the possibility of indirectly extracting the 21\,cm signal in the presence of foregrounds by taking advantage of nonlinear couplings between large and small scales, both at the level of the density field and a biased tracer of that field. When combined with effective field theory techniques, bias expansions allow for a systematic treatment of nonlinear mode-couplings in the map between the underlying density field and the 21\,cm field. We apply an effective bias expansion to density information generated with \texttt{21cmFAST} using a variety of assumptions about which information can be recovered by different surveys and density estimation techniques. We find that the 21\,cm signal produced by our effective bias expansion, combined with our least optimistic foreground assumptions, produces $\sim30\%$ cross-correlations with the true 21\,cm field. Providing complementary density information from high-redshift galaxy surveys yields a cross-correlation of 50-70\%. The techniques presented here may be combined with foreground mitigation strategies in order to improve the recovery of the large-scale 21\,cm signal.
\end{abstract}

\maketitle

\section{Introduction}
\label{sec:intro}
The time from the birth of the first stars to the end of the Epoch of Reionization (EoR) marks an important era for the astrophysics of galaxy formation and assembly, as well as for the underlying growth of cosmological structure~\cite{Loeb_Barkana2001:ARANA, Zaroubi2013:EoR_Review, Robertson2022:ARANA,McQuinn:2006et, Mesinger2016:Review, Liu2016:21cm_Neutrino}. The redshifted 21~cm line from neutral hydrogen is a direct probe of this period of cosmic time~\cite{Furlanetto:2006jb,2012RPPh...75h6901P}, which is being actively targeted by many existing and proposed radio observatories~\cite{Edges2017:Overview, SARAS3:Overview, 2019JAI.....850004P, 2010AJ....139.1468P, 2013A&A...556A...2V, NenuFAR2012:Overview, 2013PASA...30....7T, MWA2018:PhaseII_Overview, GMRT2017:uGMRT_Overview, DeBoer:2016tnn, Berkhout2024:HERA_PhaseII, LWA2019:21cmLimit, SKA2015:EoR}. In particular, radio interferometers have already set stringent upper bounds on the 21 cm power spectrum~\citep{Parsons2014:PaperLimit, Dillon2014:MWA_Limit, Dillon2015:MWA_Limit, Ewall-Wice2016:Limit_Relection, Beardsley2016:MWA_Limit, Patil2017:LofarLimit, Li2019:MWA_Limit, Barry2019:MWA_Limit, LWA2019:21cmLimit, Gehlot2019:LofarLimit, Kolopanis2019:PaperLimit, Trott2020:MWA_Limit, Mertens:2020llj, Garsden2021:LWA_Limit, Yoshiura2021:MWA_Limit, HERA2022:h1c_idr2_limit, HERA2023:h1c_idr3_limit, Wilensky2023:MWA_Limit, Munshi2024:Nenufar_Limit, Acharya2024:LOFAR_Limit, Mertens:2025pvk,Nunhokee:2025jbn, HERA:2025ajm}, which have been used to constrain the properties of the intergalactic medium (IGM) at high redshifts \cite{HERA2022:Theory_Limits,HERA2023:h1c_idr3_limit,Ghara:2025xzu,Nunhokee:2025jbn}.

A major ongoing issue in 21~cm cosmology is that the signal is highly contaminated by foreground emission that is many orders of magnitude brighter (see Ref.\,\cite{2020PASP..132f2001L} for a review). These foregrounds, which include Galactic synchrotron radiation, are expected to be spectrally smooth and will therefore affect large-scale modes of the 21\,cm signal along the line of sight. Additionally, the chromatic response of interferometers can cause the foregrounds to contaminate other Fourier modes. Decomposing each wavevector into $\mathbf{k}\equiv(\mathbf{k}_\perp, k_\parallel)$, where $k_\parallel$ denotes the Fourier modes parallel to the line of sight and $\mathbf{k}_\perp$ represents the Fourier modes in the plane perpendicular to the line of sight, then in $k_\perp$-$k_\parallel$ space, the region where foreground  contamination occurs is referred to as the ``foreground wedge'' due to its shape~\cite{2010ApJ...724..526D,2012ApJ...752..137M,2012ApJ...756..165P,Thyagarajan:2015kla,2014PhRvD..90b3018L,2014PhRvD..90b3019L,2025A&A...693A.276M}.

In principle, there is more cosmological information encoded in the wedge-obscured modes, because the scaling of the power spectrum and thermal instrumental noise is such that the signal-to-noise ratio is much higher for small wavenumbers. However,
it is difficult to completely subtract foregrounds \cite{2010ApJ...724..526D, Liu2011:QE, Chapman2012:FastICA, Ghosh2015:Bayesian_21cm, Ewall-Wice2021:DAYNENU, Mertens2018:GPR, Wang2024:Foreground_subtraction} in the wedge due to uncertainties in modeling both the foreground emission and the chromatic response of the instrument \cite{Barry2016:CalibrationError, Patil2016:CalibrationError, Ewall-Wice2017:CalibrationError, Bryne2019:CalibrationError, Mouri_Sardarabadi2019:CalibrationError, Neben2016:BeamErrorMWA, Joseph2018:BeamError, Ansah-Narh2018:BeamPerturbation, Zheng:2016kqa,Orosz2019:BeamVariation, Joseph2020:BeamVariation, Choudhuri2021:BeamVariation, Kim2022:BeamPerturbation, Ewall-Wice2016:Limit_Relection, Kern2019:Relection_Model, Ung2020:MWA_Coupling_Sim, Josaitis2022:MutualCoupling, Rath_Pascua2024:Mutual_Coupling, Chen2025:RFI}. Therefore, these modes are often discarded, with many analyses of 21~cm data focusing on evading foreground-contaminated modes~\cite{2012ApJ...745..176V,Trott:2012md, 2013ApJ...770..156H, Thyagarajan:2013eka, Liu:2014bba}. This effectively renders a large number of cosmological modes inaccessible to current experiments, significantly reducing their sensitivity. It also limits the possibility of correlating the 21~cm signal with complementary tracers of the matter field from high-redshift galaxy surveys or other line intensity mapping experiments \cite{LaPlante:2022nlp, Gagnon-Hartman2025:21cm_x_Lya, Fronenberg2024:LIM_x_21cm, Chen:2025hzj}. 

As with any biased tracer of the matter field, there is non-linear mode-coupling between large-scale and small-scale modes of the 21~cm field. In principle, it is therefore possible to take advantage of this mode-coupling to reconstruct large-scale information from uncontaminated measurements of small-scale modes. This type of reconstruction has been studied in the context of matter density and halo fields~\cite{2016PhRvD..93j3504Z,Zhu:2021qmm,Zang:2022qoj}, as well as in 21\,cm cosmology~\cite{2018PhRvD..98d3511Z,2019MNRAS.486.3864K,Darwish:2020prn}, although usually at low redshifts of $z \lesssim 5$. Past studies have also applied machine learning methods to reconstruct the foreground-obscured 21\,cm modes~\cite{Gagnon-Hartman:2021erd,Kennedy:2023zos,Sabti:2024jff,Thelie:2025jgk}. On length scales where 21\,cm fluctuations are only \textit{mildly} nonlinear, one can use techniques inspired by effective field theory (EFT) to systematically treat these nonlinearities~\cite{McQuinn:2018zwa, Qin:2022xho}. For instance, it was recently shown that EFT methods applied to small-volume (high-$k$) hydrodynamical simulations of the 21~cm field can accurately predict larger volumes~\cite{Qin:2025olv}.

\begin{figure*}[ht]
    \centering
    \includegraphics[width=\textwidth]{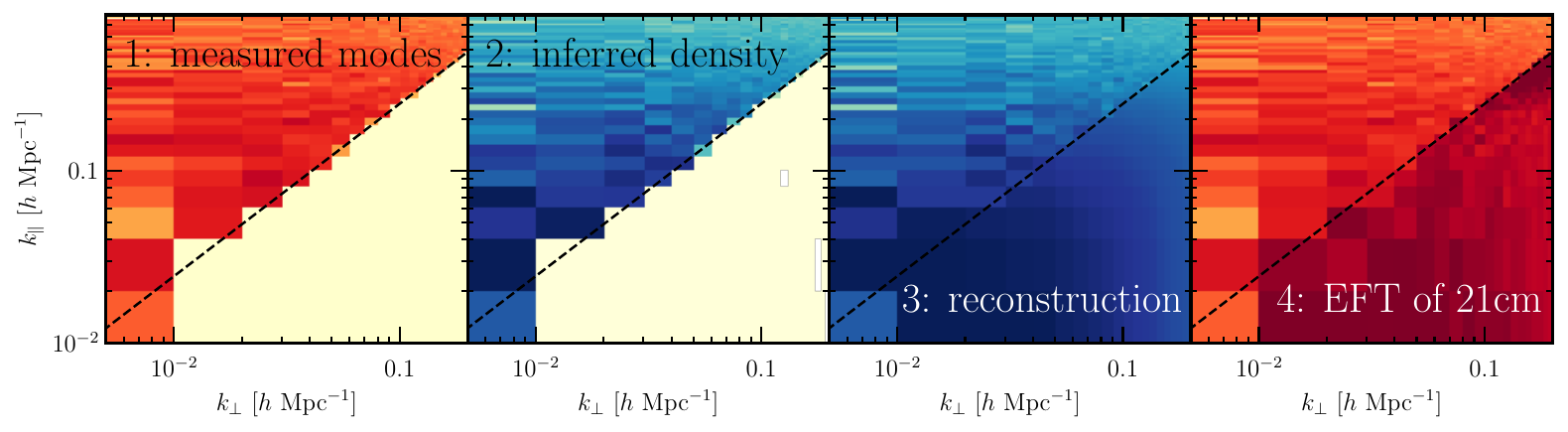}
    \vspace{-0.4cm}
    \caption{
        The steps involved in our procedure.
        Starting at the left, radio telescopes will measure high-$k$ 21\,cm modes (\textit{step 1}).
        Since the 21\,cm signal is a biased matter tracer, one can infer the underlying matter density field information, which will also mostly be at high $k_\parallel$ values (\textit{step 2}).
        We can then take advantage of nonlinear couplings to reconstruct the missing density modes (\textit{step 3}) and then use the bias expansion to translate the density fluctuations back into 21\,cm fluctuations (\textit{step 4}).
    }
    \label{fig:diagram}
\end{figure*}

In this work, we further explore the predictive power that comes from the coupling between modes with different Fourier shapes in order to see how well we can recover foreground-obscured 21~cm modes. To test the robustness of the techniques used here, we make several different assumptions about which 21~cm modes are contaminated by foregrounds and what underlying density information is available. We also assess the impact of supplemental density information inferred from other probes in different parts of Fourier space, such as galaxy surveys, e.g. by the \textit{Nancy Grace Roman Space Telescope}~\cite{2015arXiv150303757S}. 

The rest of this paper is organized as follows. In Section~\ref{sec:methods}, we provide a conceptual overview of how we reconstruct missing modes and describe the various assumptions we make about what density information is available. In Section~\ref{sec:theory}, we briefly review the effective 21\,cm bias expansion of Refs.~\cite{McQuinn:2018zwa, Qin:2022xho} and describe our methods for reconstructing missing density modes. In Section~\ref{sec:results}, we present the results of applying the bias expansion to different combinations of density information, and also demonstrate the complementarity of measurements from galaxy surveys. We conclude in Section~\ref{sec:conclusion}. Throughout this article, we use cosmological parameters from \textit{Planck} 2018~\cite{Planck:2018vyg}.

\section{Conceptual Overview}
\label{sec:methods}
The pipeline for recovering the foreground-contaminated 21\,cm signal, as schematically illustrated in Fig.~\ref{fig:diagram}, consists of four main steps:  
\begin{enumerate}
    \item Experiments such as HERA will measure the 21\,cm signal outside of the foreground wedge. 
    \item From these 21\,cm measurements, one can extract density information outside of the foreground wedge. With this density information and the 21\,cm modes from step 1, one can infer the bias parameters describing their connection.
    \item One can then reconstruct the density modes \textit{inside} the foreground region by taking advantage of nonlinear mode-coupling in the matter density field.
    \item Finally, with the bias parameters and linear density field, it is possible to map the density field back to 21\,cm field using an effective bias expansion, thus obtaining 21\,cm modes within foregrounds.
    We compare the predicted 21\,cm modes to signals generated with \texttt{21cmFAST} to validate this approach.
\end{enumerate}
While the first step poses a challenging data analysis problem, the second step in this pipeline is highly nontrivial from a theoretical perspective. In principle, since the 21\,cm radiation is a biased tracer of the density field, one should be able to express the 21\,cm field in terms of an expansion in powers of the density field, described in Section~\ref{sec:EFT}. This expansion could then be inverted in order to obtain the density fluctuations from the 21\,cm fluctuations. However, in practice, nonlinear contributions make the information relatively difficult to extract~\cite{Feng:2018for}. These challenges have been extensively studied in the context of galaxy surveys, particularly for the purpose of reconstructing the linear density field and baryon acoustic oscillations~\cite{Frisch:2001vw,Brenier:2003xs,Eisenstein:2006nk,Jasche:2012kq,Kitaura:2012tu,Tassev:2012hu,Wang:2013ep,Schmittfull:2017uhh,Seljak:2017rmr,Hada:2018fde,Modi:2018cfi,Liu:2020pvy,Zhu:2021qmm,Zang:2022qoj,Chen:2023iia}, and some limited progress has been made based on using 21\,cm information as a starting point~\cite{Modi:2019hnu,Darwish:2020prn,Zhou:2023mbb}. 

Devising an optimal strategy for extracting density information from 21\,cm fluctuations would be worth its own dedicated study, and therefore we choose to focus primarily on steps 3 and 4 in this work. However, we note that Ref.~\cite{Chen:2025wdy} has shown that step 2 is possible by adopting a gradient-based sampler to simultaneously infer the bias parameters and the underlying density field from the observed 21 cm modes outside the foreground wedge. Specifically, deep in the perturbative regime, the 21\,cm field and density field are primarily related through the linear bias term which is local~\cite{Qin:2022xho,Qin:2025olv}. This indicates that the density modes that can be recovered correspond to the same modes where one has uncontaminated 21\,cm information. 
This is corroborated by the findings of Ref.~\cite{Chen:2025wdy}, which showed explicitly that density modes outside the foreground wedge can be reliably inferred with up to $90\%$ accuracy on large scales (depending on the region of Fourier space) assuming a $\sim3\sigma$ detection of those same individual 21\,cm modes. As it turns out, the same approach of Ref.~\cite{Chen:2025wdy} can also be used to reconstruct some of the large-scale density modes in the foreground wedge because their forward model includes nonlinear mode-coupling. These results and future improvements to step 2 can interface with the framework presented here in a straightforward way.

Motivated by uncertainties about which modes can be reliably inferred, as this is an active line of inquiry, we consider multiple representative scenarios in the following discussion. Starting from the least optimistic case and moving to more ambitious possibilities, we assume that the density modes initially recovered are:
\begin{itemize}    
    \item the Fourier modes $\mathbf{k}$ outside of the foreground wedge, which are those with~\cite{Liu:2014bba}
    \begin{equation}
        |k_\parallel| \leq \frac{D_\mathrm{c}(z)H(z)}{c (1+z)} |\mathbf{k}_\perp|,
    \end{equation}
    where $D_\mathrm{c}(z)$ is the co-moving line-of-sight distance to redshift $z$ \cite{Hogg1999:Distance}, $H(z)$ is the Hubble parameter as a function of redshift, and $c$ is the speed of light. 
    We will refer to this as the ``wedge" case and it corresponds to the most likely future scenario. 

    \item the modes $k_\parallel > k_\mathrm{cut}$, corresponding to a scenario where densely spaced or specifically arranged interferometer array configurations make it possible to use foreground mitigation strategies that may be able to recover wedge modes at high $k_\perp$~\cite{2018ApJ...869...25M,MacKay:2025flx}.
    We also include this case in our discussion because it enables a more direct comparison with results from the literature~\cite{Darwish:2020prn}.
    We will refer to this as the ``flat" case.

    \item the modes $k > k_\mathrm{cut}$, which has been used in the literature to approximate the limitations of galaxy surveys~\cite{Darwish:2020prn}. We also include this case as further motivation for studies of 21~cm foreground mitigation, to show how much more of the missing modes we can recover given additional small scale information.
    We will refer to this as the ``isotropic" case.
\end{itemize}
We treat all three of these cases the same way in steps 3 and 4, with the only difference being the availability of modes entering into the reconstruction.

\section{Nonlinear mode-coupling and reconstruction}
\label{sec:theory}

\subsection{An effective bias expansion of the 21 cm signal}
\label{sec:EFT}

Here, we briefly review how one can predict the large-scale 21~cm signal from the underlying density field by applying the effective bias expansion described in Refs.~\cite{McQuinn:2018zwa, Qin:2022xho}, which corresponds to the last step in Fig.~\ref{fig:diagram}.

In linear perturbation theory, density perturbations independently evolve with scale factor $a$ as $\delta^{(1)} \sim D(a)$, regardless of the length scale. Here $D(a)$ is the linear growth factor, corresponding to $D(a) \sim a$ in a matter-dominated universe. However, as structure formation progresses, nonlinear contributions increasingly couple different Fourier modes. We define $k_\mathrm{NL}$ as the wavenumber at which nonlinear contributions become significant, i.e. $k < k_\mathrm{NL}$ or length scales greater than $\sim 1 / k_\mathrm{NL}$ can be treated perturbatively. In standard perturbation theory, the nonlinear evolution of $\delta$ is parameterized as an order-by-order expansion. Using bold subscripts to denote Fourier-transformed quantities, e.g. $(\delta)_{\mathbf{k}} \equiv \int d^3 \mathbf{x} \, \delta (\mathbf{x}) e^{-i \mathbf{k} \cdot \mathbf{x}}$, this expansion is given by
\begin{equation}
    \delta_{\boldsymbol{k}} = \sum_{n=1}^\infty a^n \delta^{(n)}_{\boldsymbol{k}},
    \label{eqn:delta_ansatz}
\end{equation}
where the higher-order terms are in turn parameterized as convolutions of the linear density field,
\begin{align}
    \delta^{(n)}_{\boldsymbol{k}} &= \int \dbar^3 \mathbf{q}_1 \dots \int \dbar^3 \mathbf{q}_n \, (2\pi)^3 \delta^D \left( \boldsymbol{k} - \sum_{i=1}^n \boldsymbol{q}_i \right) \nonumber \\ &\times F_n (\boldsymbol{q}_1, \dots, \boldsymbol{q}_n) \delta^{(1)}_{\boldsymbol{q}_1} \dots \delta^{(1)}_{\boldsymbol{q}_n}, \label{eq:SPTkernels}
\end{align}
where $\dbar^3 \mathbf{q}$ is shorthand for $d^3 \mathbf{q} / (2\pi)^3$. The Dirac delta function $\delta^D$ enforces a relationship between the coupled modes that enter into the convolution kernels $F_n$, which have well-known recursion relations~\cite{Goroff:1986ep,Jain:1993jh,Bernardeau:2001qr}. 

The matter velocity field $\mathbf{v}$ can be decomposed into its divergence, $\theta = \nabla \cdot \mathbf{v}$, and curl, $\omega = \nabla \times \mathbf{v}$. The curl component decays like $\omega \sim 1/a$ in standard perturbation theory, so we neglect it. Meanwhile, in linear perturbation theory $\theta^{(1)} \sim \delta^{(1)}$. The divergence of the velocity field can then be expressed as an expansion similar to the density field,
\begin{equation}
    \theta_{\boldsymbol{k}} = a H \sum_{n=1}^\infty a^n \theta^{(n)}_{\boldsymbol{k}},
    \label{eqn:theta_ansatz}
\end{equation}
where higher-order terms are again expressed as convolutions of the linear density field,
\begin{align}
    \theta^{(n)}_{\boldsymbol{k}} &= \int \dbar^3 q_1 \dots \int \dbar^3 q_n \, (2\pi)^3 \delta^D \left( \boldsymbol{k} - \sum_{i=1}^n \boldsymbol{q}_i \right) \nonumber \\ &\times G_n (\boldsymbol{q}_1, \dots, \boldsymbol{q}_n) \delta^{(1)}_{\boldsymbol{q}_1} \dots \delta^{(1)}_{\boldsymbol{q}_n},
    \label{eqn:theta_higher_order}
\end{align}
and where the recursion relations for the $G_n$ convolution kernels are also given in Refs.~\cite{Goroff:1986ep,Jain:1993jh,Bernardeau:2001qr}. 

We then assume that fluctuations in the 21\,cm intensity field, $\delta_{21}$, trace the nonlinear matter density, and can thus be written as a local bias expansion in terms of $\delta$. We include all operators respecting  rotational and Galilean invariance up to second order in $\delta$,
\begin{align}
    (\delta_{21})_{\boldsymbol{k}} =& b_1 \delta_{\boldsymbol{k}} - b_{\nabla^2} \left( \frac{k}{k_\mathrm{NL}} \right)^2 \delta_{\boldsymbol{k}} \n
    &\quad + b_2 \left(\delta^2\right)_{\boldsymbol{k}} + b_{\mathcal{G}2} (\mathcal{G}_2)_{\boldsymbol{k}} + \cdots
    \label{eqn:bias}
\end{align}
The tidal (Galileon) operator is defined in configuration space as
\begin{equation}
    \mathcal{G}_2 = (\nabla_i \nabla_j \phi) (\nabla^i \nabla^j \phi) - (\nabla^2 \phi)^2 ,
    \label{eqn:tidal_op}
\end{equation}
where $\phi$ is the gravitational potential. In addition to the linear and quadratic bias, the expansion in Eq.~\eqref{eqn:bias} includes $b_{\nabla^2}$, which captures local nonlinearities due to e.g. ionizing photons that travel long distances and is degenerate with leading-order EFT of LSS sound speed term~\cite{Baldauf:2014qfa}. We neglect the stochastic contribution to this expansion, as this term is negligible at the wavenumbers of interest~\cite{McQuinn:2018zwa}.

The higher-order terms appearing in Eq.~\eqref{eqn:bias} involve integrating over nonlinear scales that are not well described using perturbative methods. For example, the term $\delta^2$ which appears as a product in configuration space is written in Fourier space as a convolution,
\begin{equation}
    \left(\delta^2\right)_{\boldsymbol{k}} = \int \dbar^3 \boldsymbol{q} \, \delta_{\boldsymbol{q}} \delta_{\boldsymbol{k-q}},
\end{equation}
where the integration includes UV contributions from large wavenumbers (i.e. small length scales) which are beyond the regime of perturbative control. However, the long-wavelength modes that we are concerned with should be largely insensitive to small-scale nonlinear modes, and we thus add counterterms to each operator that cancel the UV-sensitive contributions up to a certain order in a procedure analogous to renormalization (see Refs.~\cite{Assassi:2014fva} and~\cite{Qin:2022xho} for more details). For example, the first few terms of the renormalized $\delta^2$ operator, which we denote as $[\delta^2]$, are given by
\begin{equation}
    [\delta^2] = \delta^2 - \sigma^2 (\Lambda) - \frac{68}{21} \sigma^2 (\Lambda) \delta,
\end{equation}
where $\Lambda$ is the small-scale cutoff for the perturbative theory, $\sigma^2 (\Lambda) = \int_0^\Lambda \frac{dq}{2\pi^2} \, q^2 P_\mathrm{L} (q)$ is the variance in the linear matter fluctuations for $q < \Lambda$, and $P_\mathrm{L} (q)$ is the linear matter power spectrum. The bias parameters must also shift to compensate for the newly added counterterms, and for a given parameter $b_i$, we write its renormalized counterpart as $b_i^{(R)}$.

Finally, we incorporate contributions from redshift space distortions (RSDs). Due to the peculiar velocities of 21\,cm sources, $\boldsymbol{v}_\mathrm{pec}$, the inferred coordinate of a source $\boldsymbol{x}_r$ is related to its true coordinate $\boldsymbol{x}$ by
\begin{equation}
    \boldsymbol{x}_r = \boldsymbol{x} + \frac{\hat{\boldsymbol{n}} \cdot \boldsymbol{v}_\mathrm{pec}}{a H} \hat{\boldsymbol{n}},
    \label{eqn:x_RSD}
\end{equation}
where $\hat{\boldsymbol{n}}$ is the line-of-sight direction, $a$ is the scale factor, and $H$ is the Hubble parameter. One can then express the density field in redshift space in terms of the real-space density using Eq.~\eqref{eqn:x_RSD} and obtain
\begin{align}
    (\delta_r)_{\boldsymbol{k}} = &\delta_{\boldsymbol{k}} 
    - i \frac{k_\parallel}{aH} (v_\parallel)_{\boldsymbol{k}} 
    - i \frac{k_\parallel}{aH} (\delta v_\parallel)_{\boldsymbol{k}} \n
    &- \frac12 \left( \frac{k_\parallel}{aH} \right)^2 \left( v_\parallel^2 \right)_{\boldsymbol{k}} 
    - \frac12 \left( \frac{k_\parallel}{aH} \right)^2 \left( \delta v_\parallel^2 \right)_{\boldsymbol{k}}\n
    &+ \frac{i}{6} \left( \frac{k_\parallel}{aH} \right)^3 \left( v_\parallel^3 \right)_{\boldsymbol{k}} + \cdots,
    \label{eqn:RSDs}
\end{align}
where we have defined $v_\parallel = \hat{\boldsymbol{n}} \cdot \boldsymbol{v}_\mathrm{pec}$ and $k_\parallel = \hat{\boldsymbol{n}} \cdot \boldsymbol{k}$, and Taylor expanded the expression by treating $k_\parallel v_\parallel/aH$ as a small number. This expansion is justified since, on the length scales that we are interested in, the peculiar velocities of sources are a small effect compared to the Hubble flow. The velocity fields can be calculated from the linear density field using Eqs.~\eqref{eqn:theta_ansatz} and \eqref{eqn:theta_higher_order}. This RSD expansion then contains new operators such as $v_\parallel^2$ which must also be renormalized.

Combining Eqs.~\eqref{eqn:bias} and \eqref{eqn:RSDs} and renormalizing all UV-sensitive operators, the relationship between the 21\,cm field and linear density field is thus given by
\begin{align}
    (\delta_{21,r})_{\boldsymbol{k}} = &~b_{1}^{(R)} \delta_{\boldsymbol{k}} - b_{\nabla^2} k^2 \delta_{\boldsymbol{k}} + b_{2}^{(R)} \left[ \delta^2 \right]_{\boldsymbol{k}} + b_{G2}^{(R)} (\mathcal{G}_2)_{\boldsymbol{k}} \n
    &-i \frac{k_\parallel}{aH} \left[ (v_\parallel)_{\boldsymbol{k}} + b_1 (\delta v_\parallel)_{\boldsymbol{k}} - b_{\nabla^2} k^2 (\delta v_\parallel)_{\boldsymbol{k}} \right] \n
    &- \frac{1}{2} \left(\frac{k_\parallel}{aH} \right)^2 \left[ v_\parallel^2 \right]_{\boldsymbol{k}} + \cdots
    \label{eqn:d21_renorm}
\end{align}
This model has been shown to reproduce the 21\,cm power spectrum from hydrodynamical simulations with errors at the percent level, and the field-level signal with errors at the $\mathcal{O} (10\%)$ level~\cite{Qin:2022xho}.

Since, throughout the text, we assume that both density and 21\,cm measurements are available outside of some ``wedge"-shaped, ``flat", or ``isotropic" region, we fit the renormalized bias coefficients using Eq.~\eqref{eqn:d21_renorm} only over the available wavenumbers in each scenario. We perform a field-level fit for the coefficients by minimizing the function
\begin{equation}
    \mathcal{A} = \sum_\mathbf{k} \frac{|(\delta_\mathrm{sim})_\mathbf{k} - (\delta_\mathrm{EFT})_\mathbf{k}|^2}{V},
\end{equation}
where $\delta_\mathrm{sim}$ is the 21\,cm fluctuations generated from e.g. \texttt{21cmFAST}, $\delta_\mathrm{EFT}$ is the 21\,cm signal predicted from the underlying density field using e.g. Eq.~\eqref{eqn:d21_renorm}, and $V$ is the volume of the simulation~\cite{Qin:2022xho}.

\subsection{Reconstruction of missing density modes}
\label{sec:recon}
For step 3 in Fig.~\ref{fig:diagram}, we use the procedure from Ref.~\cite{Darwish:2020prn} as one example for reconstructing missing modes at the level of the density field.
Here, we summarize the main steps for constructing the estimator; more details about how to derive the following expressions can be found in Ref.~\cite{Darwish:2020prn}. 

In linear perturbation theory, different modes of an initially Gaussian density field evolve independently of one another; however, nonlinear gravitational evolution couples modes at different length scales. 
For example, the nonlinear density field can be written in terms of the linear matter overdensity $\delta^{(1)}$ as
\begin{equation}
    (\delta_{\mathrm{NL}})_{\mathbf{k}} = \delta^{(1)}_{\mathbf{k}} + \sum_\alpha \int \dbar^3 \mathbf{q} \, F_\alpha (\mathbf{q}, \mathbf{k-q}) \delta^{(1)}_{\mathbf{q}} \delta^{(1)}_{\mathbf{q-k}} + \cdots,
    \label{eqn:expansion}
\end{equation}
where $\alpha$ runs over distinct types of quadratic couplings and $F_\alpha$ represents the corresponding kernel. The kernels we examine in this study correspond to (1) the ``growth'' coupling, which can be thought of as the enhanced or suppressed growth of a small-scale mode due to being embedded in a larger-wavelength over/underdensity,
\begin{equation}
    F_G (\mathbf{k}_1, \mathbf{k}_2) = \frac{17}{21} ;
    \label{eqn:F_G}
\end{equation}
(2) the ``shift'' coupling, which can be thought of as the displacement of a mode due to the presence of another nearby over/underdensity,
\begin{equation}
    F_S (\mathbf{k}_1, \mathbf{k}_2) = \frac12 \left( \frac{1}{k_1^2} + \frac{1}{k_2^2} \right) \mathbf{k}_1 \cdot \mathbf{k}_2 ;
\end{equation}
and (3) the ``tidal'' coupling, which can be thought of as the tidal force on an overdensity/underdensity from being embedded in a varying gravitational field,
\begin{equation}
    F_T (\mathbf{k}_1, \mathbf{k}_2) = \frac27 \left[ \frac{(\mathbf{k}_1 \cdot \mathbf{k}_2)^2}{k_1^2 k_2^2} - \frac13 \right] .
\end{equation}
At second order in standard perturbation theory, the kernel $F_2$ appearing in Eq.~\eqref{eq:SPTkernels} is the sum of all three kernels listed above~\cite{Jain:1993jh,Ma:1995ey}. These couplings also arise in configuration-space expansions describing biased tracers of matter. For example, a quadratic bias term $\delta^2$ will have a kernel corresponding to the growth coupling. Similarly, the Fourier-transformed Galileon operator defined in Eq.~\eqref{eqn:tidal_op} will have a kernel corresponding to the tidal coupling. Given density information in some part of Fourier space, we can use these mode couplings to reconstruct missing wavenumbers.

Given the nonlinear density field, the minimum variance unbiased estimator for the linear density field is given by~\cite{Darwish:2020prn}
\begin{align}
\hat{\Delta}_{\alpha, \mathbf{K}} &= N_\alpha (\mathbf{K}) \label{eqn:estimator} \\
    &\times \int \dbar^3 \mathbf{q} \, \frac{f_\alpha (\mathbf{q}, \mathbf{K}-\mathbf{q})}{2 P_\mathrm{NL} (q) P_\mathrm{NL} (|\mathbf{K} - \mathbf{q}|)} (\delta_\mathrm{NL})_\mathbf{q} (\delta_{\mathrm{NL}})_{\mathbf{K}-\mathbf{q}} \nonumber
\end{align}
where the normalization factor is given by
\begin{equation}
    N_\alpha (\mathbf{K}) = \left( \int \dbar^3 \mathbf{q} \frac{f_\alpha (\mathbf{q}, \mathbf{K} - \mathbf{q})^2}{2 P_\mathrm{NL} (q) P_\mathrm{NL} (|\mathbf{K} - \mathbf{q}|)} \right)^{-1} 
\end{equation}
and also represents noise on the estimator from cosmic variance and shot noise, $f_\alpha$ is related to the kernels described previously by
\begin{equation}
    f_\alpha (\mathbf{q}_1, \mathbf{q}_2) = 2 \left[ F_\alpha (\mathbf{q}_1 + \mathbf{q}_2, - \mathbf{q}_1) P_\mathrm{L} (q_1) + (1 \leftrightarrow 2 ) \right],
    \label{eqn:f}
\end{equation}
$P_\mathrm{NL} (q)$ is the nonlinear matter power spectrum, and $P_\mathrm{L} (q)$ the linear power spectrum. The expression in Eq.~\eqref{eqn:estimator} is derived assuming there is only a single mode-coupling, e.g. only one of the growth, shift, or tidal type couplings present. Other couplings not accounted for in the estimator become sources of noise or contamination; see Ref.~\cite{Darwish:2020prn} for details.

We have validated our calculation by checking that we can reproduce $N_\alpha (\mathbf{K})$ in Ref.~\cite{Darwish:2020prn} in the wavenumber range where our results overlap when we make the same assumptions. Moreover, in agreement with Ref.~\cite{Darwish:2020prn}, we find that the growth-type coupling has the lowest noise of the three listed above. Therefore, for the rest of this work, we will use the growth coupling when including the step of density reconstruction.

In Ref.~\cite{Darwish:2020prn}, the normalization for the estimator is derived by requiring that it be unbiased at the field level, e.g. the expectation value of the estimator is the true linear density field. As a result, the cross-spectrum of the estimator with $\delta^{(1)}$ is given by
\begin{equation}
    \langle \delta^{(1)}_\mathbf{k} \hat{\Delta}_{\alpha, \mathbf{k}'} \rangle = (2\pi)^3 \delta (\mathbf{k} - \mathbf{k}') P_\mathrm{L} (k) .
\end{equation}
This property is appropriate for applications such as cross-correlating the estimator with galaxy surveys. However, this also means that the auto-spectrum of the estimator is 
\begin{equation}
    \langle \hat{\Delta}_{\alpha, \mathbf{k}} \hat{\Delta}_{\alpha, \mathbf{k}'} \rangle = (2\pi)^3 \delta (\mathbf{k} - \mathbf{k}') N_\alpha (k) .
\end{equation}
In this work, since we are more concerned with the power spectrum of the estimator, we require that the estimator be unbiased at the level of the power spectrum. We accomplish this by replacing the normalization $N_\alpha$ in Eq.~\eqref{eqn:estimator} and instead fixing the amplitude of each Fourier mode to its value inferred from the average power spectrum so that $|\hat{\Delta}_{\alpha, \mathbf{K}}|^2 = P_\mathrm{NL} (k)$, which can be obtained from a cosmological Boltzmann code like \texttt{CAMB}~\cite{Lewis:1999bs} or \texttt{CLASS}~\cite{lesgourgues2011cosmic}. While this procedure artificially suppresses cosmic variance, it ensures that we recover the correct matter power spectrum, and subsequently the correct 21\,cm power spectrum after applying the bias expansion described in the previous Section.

Since the validity of the estimator in Eq.~\eqref{eqn:estimator} relies on the linear and quadratic terms dominating Eq.~\eqref{eqn:expansion}, the field $\delta_\mathrm{NL}$ must be smoothed using some window function $W(k)$ in order to exclude the high-$k$ modes that are the most nonlinear. The estimator in Eq.~\eqref{eqn:estimator} can then be written in terms of convolutions, and therefore calculated using the convolution theorem. For example, if we substitute Eq.~\eqref{eqn:F_G} into Eq.~\eqref{eqn:f} to obtain $f_G$ and use this to calculate the estimator, then the integral in Eq.~\eqref{eqn:estimator} can be written as a convolution over the fields
\begin{equation}
    (\delta_A)_\mathbf{k} = \frac{(\delta_\mathrm{NL}^W)_\mathbf{k}}{P_\mathrm{NL} (k)}, \quad
    (\delta_B)_\mathbf{k} = \frac{(\delta_\mathrm{NL}^W)_\mathbf{k} P_\mathrm{L} (k)}{P_\mathrm{NL} (k)} ,
    \label{eqn:da_and_db}
\end{equation}
where $\delta_\mathrm{NL}^W$ denotes the smoothed matter field. Hence, we can apply the inverse Fourier transformation to each of these fields, multiply the configuration-space fields, and Fourier transform the result to obtain the estimator. For the shift and tidal kernels, which have somewhat more complicated expressions, one can expand Eq.~\eqref{eqn:estimator} into a sum of many convolutions over quantities similar to those appearing in Eq.~\eqref{eqn:da_and_db}.

To summarize, we use the available density information (e.g. outside of foregrounds) to calculate the expressions in Eq.~\eqref{eqn:da_and_db}.
From $(\delta_A)_\mathbf{k}$ and $(\delta_B)_\mathbf{k}$ we can then calculate $\hat{\Delta}_{\alpha, \mathbf{K}}$, which provides an estimate of the missing density information.

\section{Setup and mock fields}
\label{sec:setup}
In order to calculate the right-hand side of Eq.~\eqref{eqn:d21_renorm}, one either needs knowledge of both the matter density field and velocity field, or the linear density field alone since it can be used to calculate the full density through Eq.~\eqref{eqn:delta_ansatz} and the velocity through Eq.~\eqref{eqn:theta_ansatz}. In what follows, we assume that we only have information about the underlying density and not the velocity. Therefore, we must assume knowledge of the \textit{linear} density field even if the output of step 2 includes nonlinear contributions (i.e. the full matter density field) in actuality. As we are primarily concerned with the mildly nonlinear regime, we expect any contamination from these spurious contributions to be small since it only appears at higher order.

In addition to the density information obtained under the foreground scenarios described in Section~\ref{sec:methods}, we demonstrate the impact of various assumptions about the linear density $\delta^{(1)}$:
\begin{itemize}
    \item $\delta^{(1)} = \delta_\mathrm{m, obs} \oplus 0^\text{FG}$:
    the matter density only outside of the foreground region, which we assume can be inferred from 21\,cm measurements, per step 2 in Section~\ref{sec:methods}. In the absence of information about the density, we set $\delta^{(1)} = 0$ for modes inside the foreground-contaminated region before substituting into the effective bias expansion. This shows how well we can recover the 21\,cm signal in the presence of foregrounds using \emph{only} the bias expansion, in the absence of any density reconstruction. Fourier modes of $\delta_{21}$ within the foreground region only receive nonlinear contributions in this case due to their coupling to modes outside the region. 
    
    \item $\delta^{(1)} = \delta_\mathrm{m, obs} \oplus \delta_\mathrm{m, recon}^\text{FG}$: the combination of $\delta_\mathrm{m, obs}$ and density information inside the foreground region reconstructed using the method described in Section~\ref{sec:recon}.

    \item $\delta^{(1)} = \delta_\mathrm{m, obs} \oplus \delta_\mathrm{m, gal}$: the combination of $\delta_\mathrm{m, obs}$ from 21\,cm observations and matter density fluctuations inferred from galaxy surveys, which we describe further in Sec.~\ref{sec:21cm_comparison} (including our assumptions about redshift uncertainties). For modes that are within the 21\,cm foreground region and not covered by galaxy surveys, we again set $\delta^{(1)} = 0$. 

    \item $\delta^{(1)} = \delta_\mathrm{m, full}$: the full matter density field, both inside and outside the foreground region. While this is an overly optimistic assumption, the purpose here is to show the most accurate 21\,cm signal that could in principle be recovered using our perturbative methods. 
\end{itemize}

\begin{figure*}
    \centering
    \includegraphics[width=\textwidth]{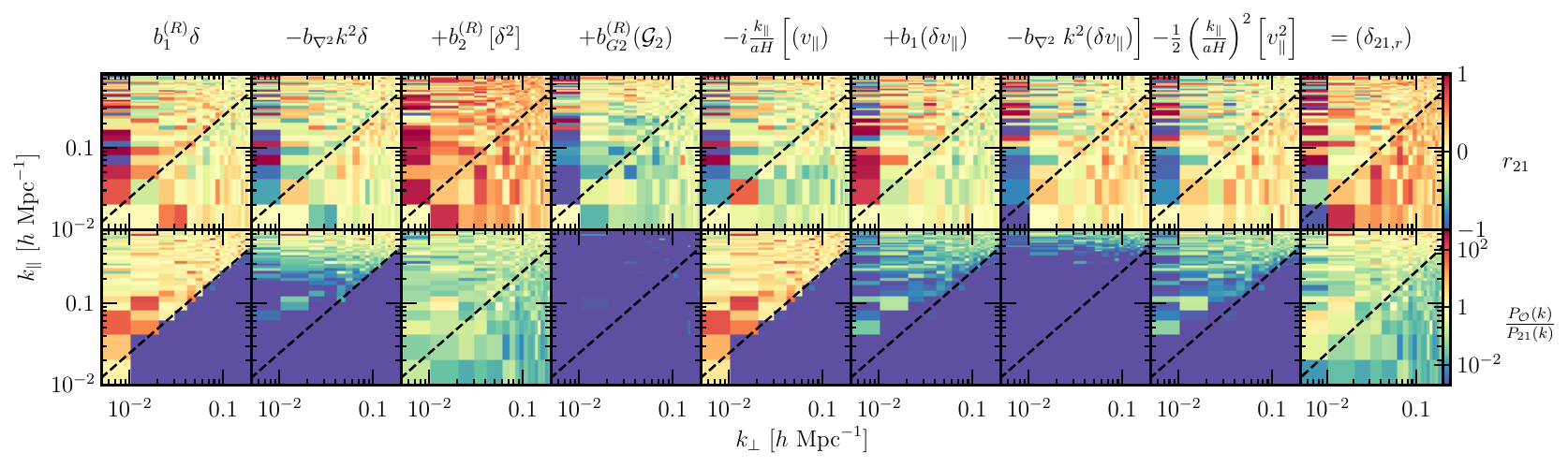}
    \vspace{-0.4cm}
    \caption{
        Terms in the bias expansion when applied to the $\delta_\mathrm{m, obs}$ with density modes missing in the foreground wedge. The first row shows the cross-correlation of each term with the true 21\,cm signal, and the second row shows the ratio of the power spectrum of each term to the 21\,cm power spectrum from \texttt{21cmFAST}. Within the foreground region, the only term with a significant amplitude is the quadratic bias ($\delta^2$), which shows a $\sim30\%$ correlation with the 21\,cm fluctuations and is the dominant contribution to the bias expansion.
    }
    \label{fig:terms_growth_EFTonly}
\end{figure*}

\begin{figure*}
    \centering
    \includegraphics[width=\textwidth]{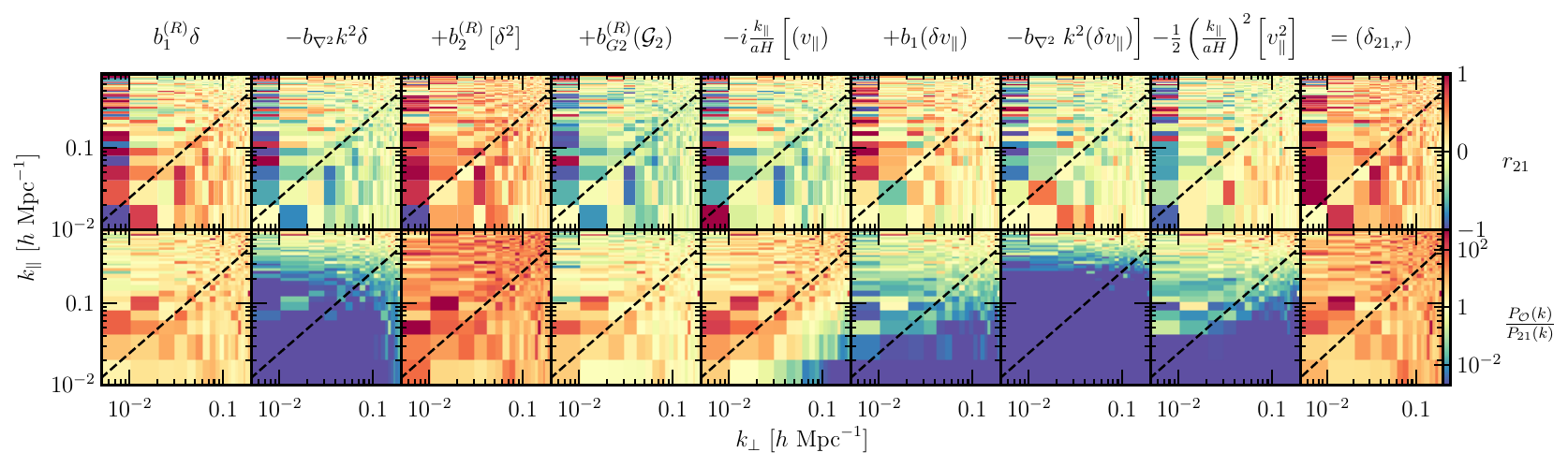}
    \vspace{-0.4cm}
    \caption{
        Same as Fig.~\ref{fig:terms_growth_EFTonly}, but applying density reconstruction to obtain density modes within the foreground wedge before mapping to 21\,cm with the bias expansion. Compared to Fig.~\ref{fig:terms_growth_EFTonly}, including density reconstruction improves the recovery of both the phase information and amplitude of the linear terms, $\delta$ and $k^2 \delta$.
        Nonlinear terms such as $\delta^2$ and $\mathcal{G}_2$ have an enhanced amplitude compared to Fig.~\ref{fig:terms_growth_EFTonly}.
        The sum of the terms, shown in the rightmost panel, has a similar cross-correlation to the true 21\,cm signal as in Fig.~\ref{fig:terms_growth_EFTonly}, since the nonlinear terms already exhibited a significant cross-correlation.
    }
    \label{fig:terms_growth_recon}
\end{figure*}

\begin{table}[t!]
  \renewcommand{\arraystretch}{1.5}
  \begin{tabular}{|c|c|c|}
    \hline
    Foregrounds & \# modes & \# perturbative \n
    \hline
    Full & 2,406,104 & 41,851 \n
    Isotropic & 2,403,899 & 39,646 \n
    Flat & 2,100,852 & 18,362 \n
    Wedge & 105,765 & 3,122 \n
    \hline
  \end{tabular} 
  \caption{
  The number of modes available from the \texttt{21cmFAST} box under each assumption about modes lost to foregrounds, as well as the number of perturbative modes ($k < k_\mathrm{NL}$) in each case.
  The nonlinear scale $k_\mathrm{NL}$ and the foreground wedge are calculated at $z=8.5$.
  }
  \label{tab:modes}
\end{table}

The mock 21~cm brightness temperature fields that we compare our methods against are generated using \texttt{21cmFAST} \cite{Mesinger2011:21cmFAST, Murray2020:21cmFAST}, a semi-numerical modeling tool capable of efficiently producing realizations of the universe during the Cosmic Dawn and the EoR. \texttt{21cmFAST} utilizes second-order Lagrangian perturbation theory (2LPT) to evolve the initial conditions of the universe to the desired redshift, and uses the excursion-set formalism \cite{Furlanetto:2004sim} to generate the process of reionization. Such a formalism assumes that reionization is driven by galaxies and uses a set of empirical scaling relations to describe the astrophysical properties of these galaxies. In this work, we assume a saturated spin temperature and ignore the effect of X-ray heating. We adopt the astrophysical parametrization and fiducial parameters described in Ref.~\cite{Park:2019NewParamtrization} to model reionization, which was shown to describe high-redshift galaxy luminosity functions well (up to $z \sim 10$). For the purpose of testing our reconstruction methods, we generate a $(500\,\mathrm{cMpc})^3$ box on a $134^3$ grid at a redshift of $z=8.5$.

We use \texttt{21cmFAST} because of its ability to quickly generate such large volumes, which is necessary for testing the reconstruction of large-scale modes. The methods studied in this work could also be tested against other analytic and semi-analytic methods that can generate large volumes, such as \texttt{Zeus21}~\cite{Munoz:2023kkg}. In this work, we do not validate our method against hydrodynamical simulations, which in principle capture a richer description of the astrophysical complexities underlying reionization. However, if such large hydrodynamical simulations were available, we expect that our methods would also work well in that context, since the effective bias expansion is able to accurately describe nonlinear mode-coupling for the purpose of enlarging simulations~\cite{Qin:2022xho,Qin:2025olv}.

After generating the fields with \texttt{21cmFAST}, we remove modes that we assume are contaminated by foregrounds using the cuts described in Section~\ref{sec:recon}. For the ``flat" and ``isotropic" foregrounds, we use $k_\mathrm{cut} = 0.15$ Mpc$^{-1}$. The number of modes left uncontaminated by foregrounds is given in the second column of Table~\ref{tab:modes}; the maximum number of modes is $134^3 = 2,406,104$. 

To evaluate the expression in Eq.~\eqref{eqn:estimator}, we first filter the nonlinear small scales from the tracer field. As above, we denote the smoothed tracer field by $\delta_\mathrm{NL}^W$. We calculate the wavenumber cutoff $k_\mathrm{NL}$ for this filter by smoothing $\delta_\mathrm{NL}$ with a Gaussian window function on increasingly large length scales until the largest overdensity is less than 0.8; the value of $k_\mathrm{NL}$ is not particularly sensitive to the choice of value that we use for largest overdensity or the choice of window function~\cite{Qin:2022xho}. For the \texttt{21cmFAST} realization used in this work, we find that the nonlinear scale is approximately $k_\mathrm{NL} = 0.4$ cMpc$^{-1}$ at z=8.5. The third column of Table~\ref{tab:modes} shows the number of uncontaminated modes that are in the perturbative regime, i.e. with $k < k_\mathrm{NL}$. We note that we have chosen the resolution of our \texttt{21cmFAST} simulation to sample up to a scale that is twice as large as the nonlinear scale (i.e., $k_\mathrm{max}\sim 0.8\ \mathrm{cMpc^{-1}}$ for our simulation). To evaluate the total power spectra in Eq.~\eqref{eqn:da_and_db}, we use the \texttt{CAMB} Boltzmann code~\cite{2011ascl.soft02026L}.

\begin{figure*}
    \centering
    \includegraphics[width=\textwidth]{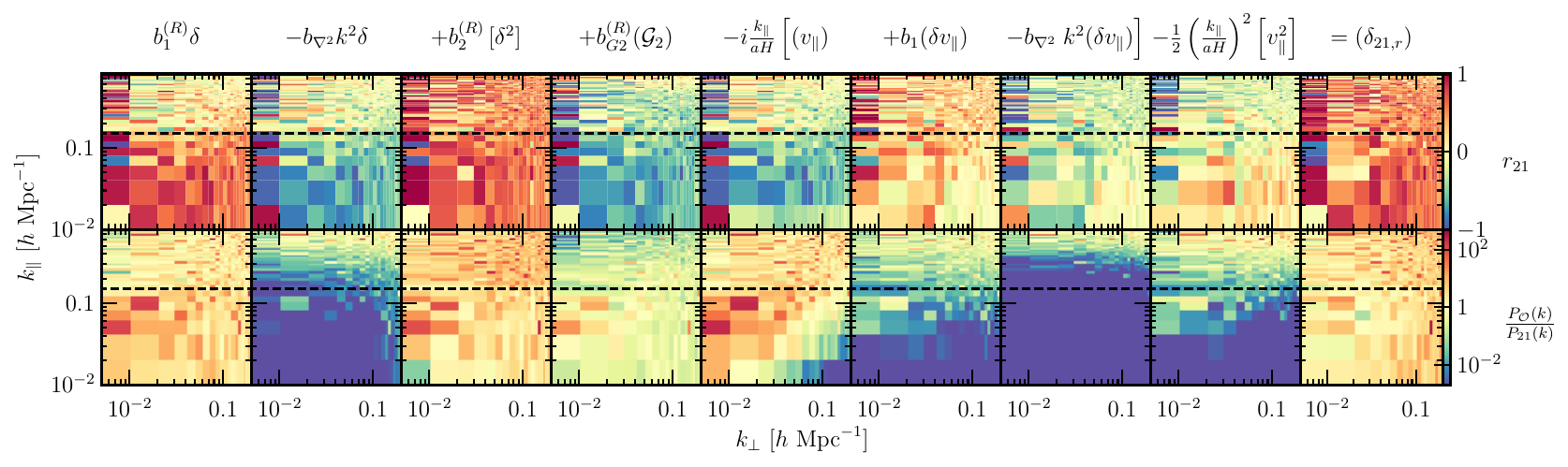}
    \includegraphics[width=\textwidth]{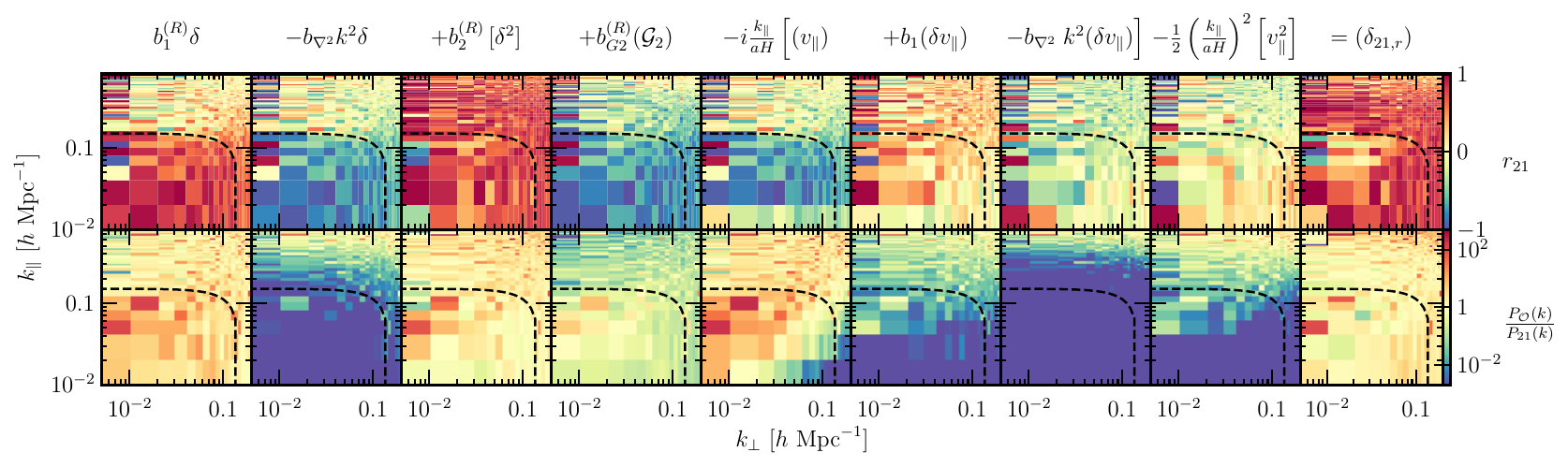}
    \caption{
        Same as Fig.~\ref{fig:terms_growth_recon}, but now assuming the flat (\textit{top panel}) and isotropic (\textit{bottom panel}) foreground shapes. Both cases show improved accuracy in the cross-correlation and power spectrum in foreground-contaminated regions as compared to the case with wedge-shaped foregrounds shown in Fig.~\ref{fig:terms_growth_recon}.
    }
    \label{fig:terms_other_FG}
\end{figure*}

\section{Results}
\label{sec:results}

In this Section, we apply the methods described in Section~\ref{sec:methods} to fields generated using \texttt{21cmFAST}. We use two metrics for determining the success of recovering the 21\,cm signal. The first is the cross-correlation between a field and the true 21\,cm signal, which is given by
\begin{equation}
    r_{21} (\mathbf{k}) = \frac{P_{X, 21} (\mathbf{k})}{\sqrt{P_X (\mathbf{k}) P_{21} (\mathbf{k})}} ,
\end{equation}
where $X$ represents any field. Note from the normalization of this expression that $r_{21}$ is insensitive to the amplitude of the field $X$ and therefore only encodes information about how well the phase of each Fourier mode in $X$ matches the 21\,cm field realization. To capture the amplitude information, we can instead look at the ratio of the power spectrum of the desired field, which does not contain phase information, to the power spectrum of the mock 21\,cm signal, $P_X (\mathbf{k}) / P_{21} (\mathbf{k})$. 

\subsection{Individual terms}
\label{sec:terms}
We first compare individual terms in the bias expansion to the mock 21\,cm signal to determine which terms hold the greatest importance for an accurate reconstruction. We compute both $r_{21}$ and $P_X/P_{21}$ for each of the eight individual terms appearing in Eq.~\eqref{eqn:d21_renorm}, including the best-fit bias coefficients and any other prefactors that multiply the RSD terms. We can then compare these to the values of $r_{21}$ and $P_X/P_{21}$ obtained by taking the sum of all the terms in the bias expansion. Note that we take this sum before computing the power spectrum, since the power spectrum of a sum is not equal to the sum of individual power spectra.

Fig.~\ref{fig:terms_growth_EFTonly} shows the result of applying the bias expansion assuming $\delta^{(1)} = \delta_\mathrm{m, obs}$ outside of the foreground wedge and $\delta^{(1)} =0$ inside the wedge, i.e. without performing any density reconstruction. The only terms that show any significant amount of cross-correlation with the true 21\,cm field in the wedge are those corresponding to $\delta^2$ and $\mathcal{G}_2$, with typical $r_{21}$ values of $0.3$ and $-0.2$, respectively. Moreover, $\delta^2$ is the only term with a large enough amplitude inside the wedge to contribute significantly to the overall bias expansion, with all other terms comprising sub-percent level contributions to the power spectrum in the foreground wedge. It is perhaps surprising that there is any appreciable level of correlation with the 21~cm field inside the wedge given that we are assuming that the initial linear density is identically zero in that region of Fourier space. Terms in Eq.~\eqref{eqn:d21_renorm} that are linear in $\delta^{(1)}$, which lack mode coupling, are therefore greatly suppressed relative to nonlinear terms such as $\delta^2$ and $\mathcal{G}_2$, although we note that even $\delta = \delta^{(1)}+\delta^{(2)}+\ldots$ has nonlinear contributions appearing in the linear bias. Nevertheless, these nonlinear terms by themselves do have some predictive power in the foreground wedge.

Applying the estimator described in Section~\ref{sec:recon} to reconstruct density information within foregrounds (i.e. substituting $\delta_\mathrm{m,obs} \oplus \delta_\mathrm{m,recon}^\text{FG}$ into Eq.~\eqref{eqn:d21_renorm}), we obtain the term-by-term cross correlations and power spectra shown in Fig.~\ref{fig:terms_growth_recon}. The linear terms in the expansion now show improved cross-correlation in the wedge at the level of about $\pm 0.2$, and also exhibit an increase in the amplitude of their power spectra. Fully nonlinear terms like $\delta^2$ and $\mathcal{G}_2$ show a similar level of cross-correlation as in Fig.~\ref{fig:terms_growth_EFTonly}; however, their amplitude in the foreground wedge is now magnified, with $P_{\mathcal{G}_2}$ exceeding the true power spectrum by a factor of about 3.5 and $P_{\delta^2}$ exceeding the true power spectrum by a factor of about 35. Although we see an improvement in the signal recovery on a term-by-term basis (particularly in the linear terms) once we include reconstructed density modes, we find that the improvement is less significant with all the terms summed together. This is because the most-improved terms are intrinsically less correlated to the true 21\,cm field, and thus their better reconstruction does not significantly improve overall accuracy.

In Fig.~\ref{fig:terms_other_FG}, we perform a similar analysis assuming different foreground shapes prior to performing density reconstruction. In these cases, we see that the cross-correlation across the first several terms is higher than for the wedge-shaped foreground region: for the ``flat" foregrounds, the first five terms all show cross-correlations of $|r_{21}| \sim 0.5$--$0.6$, and for the ``isotropic" foregrounds, we see cross-correlations of $|r_{21}| \sim 0.6$--$0.7$. Since the ``flat" and ``isotropic" foreground shapes include more modes in the perturbative regime, compared to the wedge-shaped case (see again Table~\ref{tab:modes}), this leads to an improved reconstruction of modes within the foreground region, and therefore the cross-correlation for the linear terms is now comparable to that of the nonlinear terms. This is also evident in the fact that the ``isotropic'' scenario contains more perturbative modes than the ``flat'' case, and therefore shows improved cross-correlation over the flat foregrounds.

Summing all the individual terms to obtain the total 21\,cm field, we find that the cross-correlation in the foreground region is higher than for the wedge-shaped foregrounds, and the recovered amplitude of the power spectrum is closer to that of the \texttt{21cmFAST} power spectrum. While the recovered power spectrum in the presence of wedge-shaped foregrounds is too large by a factor of about 23, this improves to only a factor of 6.4 for flat-shaped foregrounds and 4.0 for isotropic foregrounds. This improvement can again be attributed to the fact that these foreground shapes mean that our procedure starts with more perturbative modes, and thus we have better information with which to accurately reconstruct the amplitude of modes in the foreground region, as well as calculate the contribution of nonlinear terms.

\begin{figure*}
    \centering
    \hspace{-12mm}\includegraphics[width=0.9\textwidth]{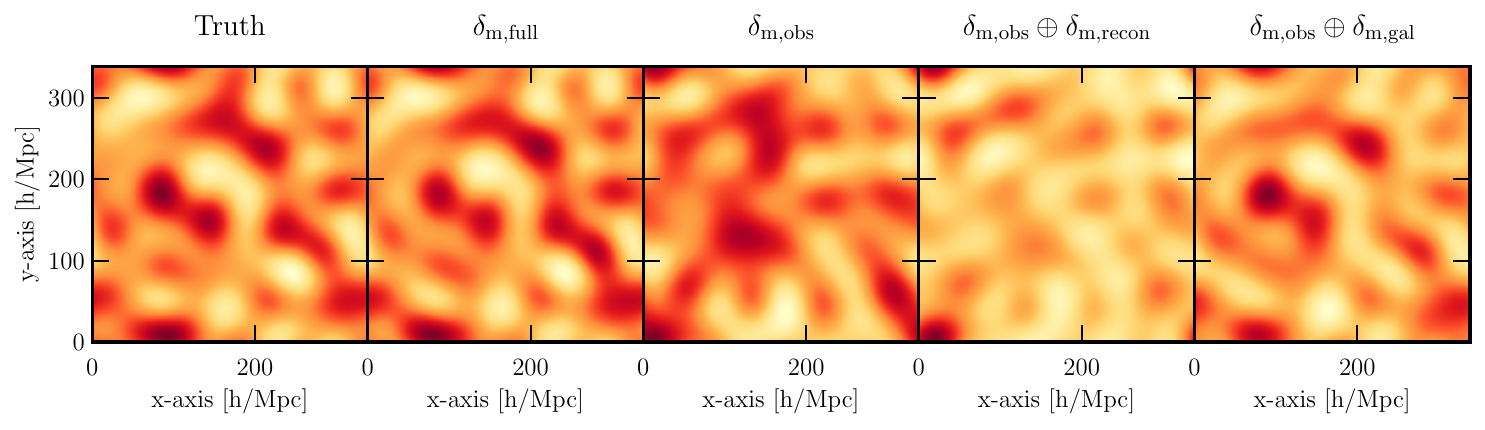}\
    \caption{
        The 21\,cm intensity field resulting from different assumptions about the density field substituted into Eq.~\eqref{eqn:d21_renorm} and smoothed over $k > 0.1$ cMpc$^{-1}$.
        In order to facilitate comparison between the different input density fields, we consistently use the bias coefficients fit using the full density field and normalize each panel arbitrarily.
    }
    \label{fig:2d_slices}
\end{figure*}

\begin{figure*}
    \centering
    \includegraphics[width=0.8\textwidth]{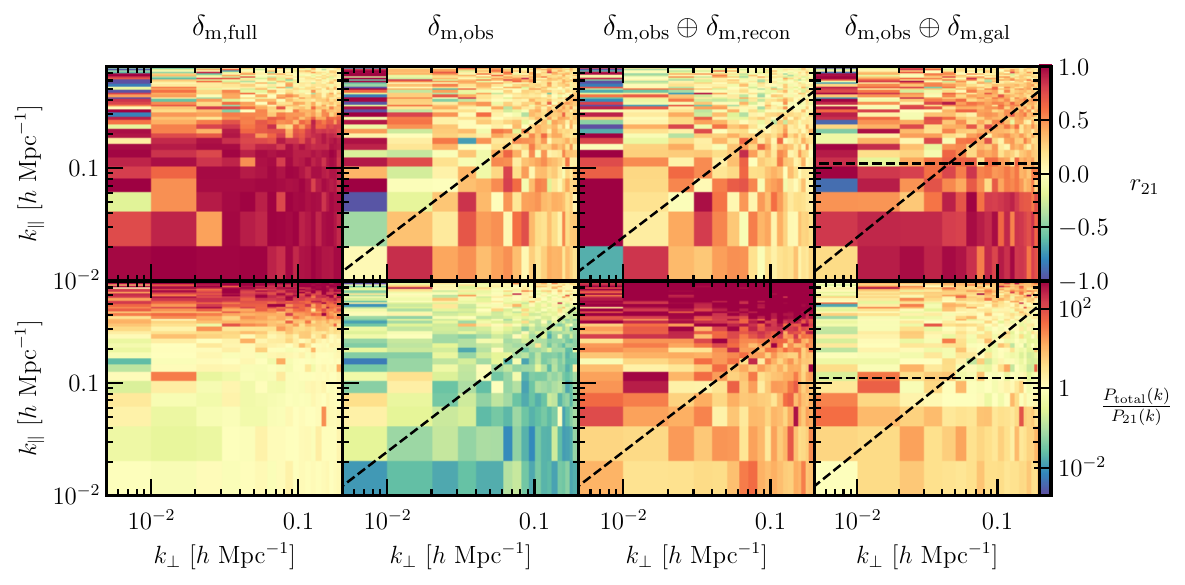}
    \caption{
        Metrics of success for reconstructing the 21\,cm signal.
        The first row shows the cross-correlation with the true 21\,cm signal, $r_\mathrm{21}$, and the second row shows the ratio of the power spectra of the two fields.
        For the last three columns, we assume $\delta_\mathrm{m,obs}$ are the density modes lying outside of a wedge-shaped foreground region.
        In order to facilitate comparison of the different input density fields, we use the bias coefficients fit using the full density field across all columns.
    }
    \label{fig:total}
\end{figure*}

\subsection{Adding density information from galaxy surveys}
\label{sec:21cm_comparison}
Galaxies are a biased tracer of the underlying matter density, and a large number of galaxies present during the EoR will soon be observed thanks to ground and space-based surveys \cite{2015ApJ...803...34B, 2015ApJ...810...71F, 2017ApJ...837...11B, 2023ApJS..269...16R, 2023ApJ...951L..22A}. However, the density field inferred from galaxies is only accurate for wavenumbers less than some value of $k_\parallel$, as uncertainties in the measured redshifts of the observed galaxies $\sigma_z$ translate to uncertainties in their line-of-sight distances. Here, we assume a galaxy survey can measure the density field below $k_\parallel \leq H(z) / c /\sigma_z$~\cite{LaPlante:2022nlp}. We note that as galaxy surveys have high angular resolution, this $k_\parallel$ threshold is constant across $k_\perp$. Typically, photometric redshift uncertainties are around $\sigma_z\sim0.5$ \cite{2015ApJ...803...34B, 2016PASA...33...37F} for dropout galaxies and $\sigma_z\sim0.1$ for galaxies selected with a narrow-band filter \cite{2017ApJ...842L..22Z, 2023ApJS..268...24K}. With follow-up observations by spectroscopic instruments, these measurements can be significantly improved with uncertainties as low as e.g. $\sigma_z \sim 0.001$ for the grism spectroscopy that will be available for the High-Latitude Wide Area Survey from the \textit{Nancy Grace Roman Space Telescope} \cite{2015arXiv150303757S, 2025arXiv250510574O}. In this discussion, we will be conservative by assuming a relatively large redshift uncertainty of $\sigma_z \sim 0.05$, with the understanding that improvements to this will only increase the level of signal recovery. This redshift uncertainty corresponds to measuring $k_\parallel < 0.11 h$~Mpc$^{-1}$ in the density field.

In Fig.~\ref{fig:2d_slices}, we show the 21\,cm intensity field calculated assuming wedge-shaped foregrounds using different assumptions about the underlying density field and smoothed over $k > 0.1$ cMpc$^{-1}$ (which is on the same order as $k_\mathrm{cut}$).
In order to ensure that differences between each panel are not merely due to changes in the best-fit bias parameters, in each panel we use the same bias parameters as determined from the full density field.
Each panel also has an arbitrary normalization in order to facilitate visual comparison of features.
The first panel shows the ``true'' \texttt{21cmFAST} field that we are attempting to reconstruct.
The second panel shows the result of substituting the full density field, $\delta_\mathrm{m,full}$, into Eq.~\eqref{eqn:d21_renorm}. 
This is the best recovery of the 21\,cm signal that one could ever obtain using the bias expansion, and there is indeed a very good resemblance to the true signal. 
The third and fourth columns serve as a comparison of the total reconstructed 21~cm field in the wedge before and after performing density reconstruction in the wedge and adding this information to observed modes. 
The phases of the 21~cm modes are not as well reconstructed upon performing density reconstruction in the wedge, hence we do not see the same features as in the first and second panels.
The last column of Fig.~\ref{fig:total} shows the result of substituting $\delta_\mathrm{m,obs} \oplus \delta_\mathrm{m,Roman}$ into Eq.~\eqref{eqn:d21_renorm}.
In this case, the reconstruction is good enough that we again see prominent 21-cm intensity features in the same regions as in the first panel.

In Fig.~\ref{fig:total}, we quantify the level of similarity between the true and reconstructed signals, and show $r_{21}$ and $P_\mathrm{total} / P_{21}$ for the same sets of assumptions as in Fig.~\ref{fig:2d_slices}. In the first column, we see that the cross-correlation is about 0.9 for wavenumbers within the perturbative regime. The amplitude of the power spectrum is also recovered with high accuracy and is within a factor of 1.5 of the true 21\,cm power spectrum.

Since the middle two columns show results for the total reconstructed 21~cm field in the wedge with and without performing density reconstruction in the wedge, they are quite similar to the final columns of Figs.~\ref{fig:terms_growth_EFTonly} and Figs.~\ref{fig:terms_growth_recon}, but using a slightly different set of bias parameters. As discussed in Section~\ref{sec:terms}, the cross-correlation is slightly better \emph{without} first performing density reconstruction. In other words, even if we assume that the matter density in the foreground wedge is \emph{zero}, which is clearly incorrect, the mode-coupling intrinsic to the EFT terms is enough to yield a cross-correlation at the level of $r\sim 0.3$. This behavior can be understood as arising from the relative contribution of the linear terms with and without in-wedge density reconstruction, as described in Section~\ref{sec:terms}: the slight degradation of the cross-correlation is due to the fact that the linear terms in the bias expansion are less correlated with the true 21\,cm fluctuations than the nonlinear terms. Additionally, without first performing density reconstruction, the reconstructed in-wedge 21~cm field underestimates the power spectrum by a factor of 8, due to the fact that we are missing contributions to the 21\,cm fluctuations from many of the underlying density modes, whereas with density reconstruction the reconstructed 21~cm field overestimates the power spectrum by a factor of 90 (this is larger than what was reported in the previous section due to the different bias coefficients).

In the last column, for modes that are within the foreground wedge and also not covered by the galaxy survey, the recovered cross-correlation is at the level of 0.5, and the power spectrum is within a factor of a few of the true amplitude. This is a clear improvement over using either $\delta_\mathrm{m,obs}$ or $\delta_\mathrm{m,obs} \oplus \delta_\mathrm{m,recon}$. This highlights the complementarity of high redshift galaxy surveys to 21cm experiments, and motivates their joint use in recovering the underlying density field as well as reconstruction of 21\,cm modes obscured by foregrounds.

\section{Conclusion}
\label{sec:conclusion}
Separating the information in long-wavelength modes along the line of sight from foregrounds is a longstanding problem of 21\,cm cosmology. Bias expansions provide a map between density fluctuations and tracers such as 21\,cm fluctuations, and may thus be used to indirectly recover 21\,cm information within foregrounds. In this work, we have tested how well one can reconstruct the 21\,cm fluctuations using the bias expansion under different scenarios about the available density information.

Assuming that we have a method of obtaining density information outside of the foreground wedge \cite{Chen:2025wdy}, we find that simply applying our bias expansion to these modes already yields $\sim 30\%$ cross-correlations with the true 21\,cm signal within the foreground region, due to non-linear couplings between small- and large-scale modes. Applying the density reconstruction method of Ref.~\cite{Darwish:2020prn} to obtain additional density information \textit{within} the foreground wedge improves the recovery of linear contributions to the 21\,cm signal within the foreground region. 
For the wedge-shaped foreground scenario, density reconstruction does not improve the overall cross-correlation. 
Density reconstruction applied to the ``flat" and ``isotropic" scenarios yields results that are much improved compared to the density-reconstructed ``wedge" case, giving cross-correlations as high as $~70 \%$ and power spectra that are within a factor of a few of the true value---this motivates further study of independent methods for extracting small-scale density information.

For comparison, we also study the impact of including density information from galaxy surveys. This information is very complementary, as galaxy surveys tend to be sensitive to low values of $k_\parallel$ less than some value corresponding to redshift uncertainties, which are the most contaminated modes in 21\,cm observations. We find that with the inclusion of galaxy survey information, the cross-correlation of recovered 21\,cm modes with the true 21\,cm signal can be as high as $50\%$ even for our least optimistic foreground assumptions.

From an information theory point of view, the reconstructed large-scale 21 cm modes do not add additional information at the field level, as they are determined solely from the small-scale modes. However, our results nonetheless demonstrate that the EFT is capable of capturing mode couplings generated by astrophysics-driven modeling frameworks such as \texttt{21cmFAST}. Moreover, these reconstructed large-scale modes enable straightforward cross-correlation with other large-scale probes such as galaxy surveys without resorting to higher-order statistics such as the bispectrum. Our method also provides a potential route to create large-scale 21 cm templates at the field level to uncover foreground-contaminated modes in the data through a matched-filter approach \cite{Wang2024:Foreground_subtraction}.

We emphasize that the methods presented here can build upon density reconstruction methods such as those presented in Ref.~\cite{Darwish:2020prn}, but are conceptually different. While that work used measurements of biased tracers to reconstruct the matter density on scales obscured by foregrounds, the present work is concerned with whether one can use information about the density in small scales to obtain the 21\,cm modes obscured by foregrounds. In this work, we have only explored the interface of the two methods by using the density estimator of Ref.~\cite{Darwish:2020prn} to supplement the small-scale density information assumed here. However, it may be instructive in the future to explore other ways in which these methods complement each other.

In summary, we have demonstrated that bias expansions combined with EFT-based treatments of nonlinear mode-coupling provide a useful starting point for recovering 21\,cm information lost to foregrounds. Their utility is enhanced when combined with reliable extractions of the underlying density field within the foreground region, e.g. from galaxy surveys. There are a number of potential directions for improvement on the results presented here. For example, the density estimator of Ref.~\cite{Darwish:2020prn} that we apply throughout is derived using an unrenormalized bias expansion; it would be instructive to rederive this estimator using renormalized operators and including all of the appropriate counterterms. Moreover, an assumption that we make throughout this work is that one can extract density information outside of the 21\,cm foreground wedge from 21\,cm measurements due to the strength of the linear bias term. However, this assumption merits a detailed study in its own right and new techniques to extract this density information, such as those presented in Ref.~\cite{Chen:2025wdy}, could be combined with the techniques studied here in a straightforward way. We leave these as directions for future study.

\section*{Acknowledgments}
It is a pleasure to thank Simon Foreman and Julian Mu\~noz for helpful discussions and comments on this paper. W.Q. was supported by the National Science Foundation Graduate Research Fellowship under Grant No. 2141064 and a grant from the Simons Foundation (Grant Number SFI-MPS-SFJ-00006250, W.Q.). K.-F.C. acknowledges support from the Mitacs Globalink Research Award, Taiwan Think Global Education Trust Scholarship, and the Taiwan Ministry of Education’s Government Scholarship to Study Abroad. K.S. acknowledges support from a Natural Sciences and Engineering Research Council of Canada (NSERC) Subatomic Physics Discovery Grant and from the Canada Research Chairs program. A.L. acknowledges support from an NSERC Discovery Grant and a Discovery Launch Supplement, and the William Dawson Scholarship at McGill. This analysis made use of \texttt{Numpy} \cite{Harris:2020xlr}, \texttt{Scipy} \cite{2020NatMe..17..261V}, \texttt{Jupyter}~\cite{Kluyver2016JupyterN}, \texttt{tqdm}~\cite{daCosta-Luis2019}, and \texttt{Matplotlib} \cite{Hunter:2007ouj}.

\bibliography{refs}
\end{document}